\documentstyle[12pt,epsf]{article}
\begin{document}
\large
\begin{titlepage}
\vspace*{4cm}
\begin{center} 
\large
Parameters of Fermi-motion from Quasielastic Backward Pion-Proton
Scattering  on Nuclei.
\end{center}
\begin{center}
 B.M.Abramov, S.A.Bulychjov, I.A.Dukhovskoy, A.I.Khanov, Y.S.Krestnikov, A.P.Krutenkova, V.V.Kulikov, M.A.Matsuk, I.A.Radkevich, E.N.Turdakina \\[50pt]
\end{center}
\begin{center}
 Abstract
\end{center}
\par
In experiment on the study of the quasielastic pion-proton scattering at
 large momentum transfer on nuclei the proton Fermi-momentum distributions
 have been analysed in plane-wave approximation for light nuclei $^6$Li, $^7$Li and $^{12}$C.
It was found that, contrary to  (e,e') experiments,
 the oscillator model gives slightly better description of our data than
 the Fermi - gas model.
But the values of parameters of the distributions obtained in our analysis
 are considerably smaller, than in (e, e') experiments. It gives evidence
 that the plane-wave approximation is not sufficient and more complicated
 theoretical models which take  into account the effects of distortion of
 pion-nucleon  amplitude in a nuclear medium are  necessary for analysis of
 our data.\end{titlepage}
\par
 Introduction
\vspace*{0.5cm}
\par
  In the last years  an interest to a problem of modification of the elementary
 particle interaction in a nuclear medium increased considerably. It is
 connected partly with development of the new theoretical approach [1],
 where the  so-called " collision broadening "  i.e. broadening of baryon
 resonances in a nuclear medium due to collisions of a resonance with nucleons
 of a nucleus is taken into account. The validity of this approach is confirmed
 by the analysis of new data on the total interaction cross sections of tagged
 photons with nuclei. For other interactions such checks have not been done yet.
 Backward quasielastic pion-nucleon scattering on nuclei at resonance  energies
 is  perspective for searches of the effects of modification of the interaction
 amplitude  in a nuclear medium. It is connected with existence  of the  baryon
 resonances and their interference with non-resonant background, which, for
 large scattering angles, leads  to strong dependence of the differential
 cross section of a pion-nucleon scattering on kinematic  variables in the
  resonance region. We have performed an experiment on systematic study of
 a backward quasielastic pion-nucleon scattering  on  nuclei at the ITEP
 accelerator. The results obtained in the framework of the  given experiment
 could be important for check of the models of pion interaction with the
 nucleon bound in nucleus, for example, those which take into account a
 non-nucleon components in the wave functions of the  bound nucleon.
 Experimental set-up will be briefly described below, and the first results
 of the analysis of the Fermi-momentum distributions for light nuclei will
 be outlined.
\par
\vspace*{0.5cm}
  Experimental procedure
\vspace*{0.5cm}
\par
The experiment was performed using the three-meter magnetic spectrometer
 (field volume 3x1x0,5 m$^3$ ) with spark chambers.
The nuclear targets were placed in a centre of a magnet. The momenta of
 the reaction products (forward scattered proton  and backward scattered
 pion ) were measured in spark chambers with high accuracy ( about 10 MeV/c)
 as well as a beam pion.  The protons were identified by time-of-flight. 
 The data were obtained at three values of momentum of a projectile pion
 p$_0 =$ 0.7, 0.9 and 1.25 GeV/c  in a range of pion-nucleon scattering
 angles from 145 up to 180 degrees in c.m.s. on the $H_2O$, $D_2O$, $^6Li$,
 $^7Li$, $^{12}C$, $^{27}Al$, $^{63}Cu$, $^{115}In$, $^{181}Ta$, $^{209}Bi$,
 $^{181}Ta$ and  $^{209}Bi$ targets.  In spite of small value of backward
 scattering cross section, the use of a wide-aperture magnetic spectrometer
 and high-intensity pion beam from ITEP accelerator  allowed to obtain large
 amount of information ( about 0.5 millions triggers ).
 An important feature of the experiment was a  registration  of the reaction
 on free protons, contained in water target.
 The scheme of the set up  was close to that shown in [2].
 As a result of the experiment the angular and the momentum distributions
 of protons knocked out forward and of pions scattered backwards were obtained.
\par
\vspace*{0.5cm}
Results
\vspace*{0.5cm}
\par
 The differential cross section of a quasielastic pion-proton scattering is a
 function of five variables ( momentum and angle of produced proton and
 momentum and two angles of a scattered pion ), therefore the analysis of these
 data is reasonable  only in the framework of models. The first step in the
 analysis of data is an adequate choice of the models and of their parameters.
 We have performed the analysis of the data in the plane-wave approximation,
 where all observables  are expressed in terms of spectral function of nuclei
 and differential cross sections of the elastic pion-proton scattering on a
 free nucleon. In our experiment the momenta of  incident pion, of the proton
 knocked out forward and of the  backward scattered  pion are measured. 
 So it is possible to calculate for each event the excitation energy of
 a residual nucleus and vector-momentum of an intranuclear nucleon, on which
 the scattering takes place. A peak of quasielastic scattering with width of
 30-40 MeV ( for $^6Li$, $^7Li$ and  $^{12}C$) dominates excitation energy
 distributions for the residual nucleus.
 The resolution in excitation energy ($\sigma =$ 12 MeV) was extracted from
 the data obtained on a hydrogen contained in a water target.
 The distributions on the probability $W(k)$ to observe a definite value
 of a module  of Fermi - momentum $k$ ($W(k)dk = \epsilon(k)|\phi(k)|^2k^2dk$,
 where $\phi(k)$ is wave function in momentum space, $\epsilon(k)$ is the
 efficiency of our  apparatus ) are shown in Fig. 1 for $^2D$, $^6Li$, $^7Li$
 and $^{12}C$ nuclei, correspondingly , at p$_0 =$ 0.7 GeV/c.  Solid curve
 in Fig. 1 for $^2D$  shows the calculated distribution for Hulten wave
 function. For other nuclei the best fits to our data in  Fermi - gas model
 (dotted curve) and in oscillator model (solid curve) are displayed. The
 parameters of  the best fits are presented in Tab.1, together with parameters,
 received in ( e, e')  experiments [3].
\par
\vspace*{0.5cm}
Conclusion
\vspace*{0.5cm}
\par
 For deuterium, the parameter  $\beta$  of Hulten distribution  found here
 agrees well with the standard value. For other nuclei, contrary to (e, e')
 experiments, the  Fermi-gas model gives slightly worse description
 of our data, than the oscillator model. Besides, the calculated parameters
 of distributions for both models are essentially smaller, than those obtained
 from (e, e') experiments. Such difference is probably due to the 
 periphericity of the  quasielastic pion-proton scattering . It is possible
 to expect, that more complicated  models such as the distorted wave impulse
 approximation  model, which  takes into account the correlation of a
 Fermi-momentum distributions with local density of a nucleus and absorption
 of pions and protons in nuclear medium, will give better description in
 comparison with the plane wave approximation used here. An  analysis of an
 energy dependence of the quasielastic pion-proton scattering in framework of
 such models can also give an information on medium modification of
 pion-nucleon scattering amplitude . 
\par
The authors are  grateful to Prof. L.A.Kondratyuk for useful discussions.
 The present work is performed under partial support of a Russian Fund for
 Basic Research, grant 96-02-17617.
\par
\newpage
\vspace*{0.5cm}
References.
\vspace*{0.5cm}
\par
1. N.Bianchi, E.De Sanctis, V.Muccifora, L.A. Kondratyuk \\ and M.I. Krivoruchenko.
 Nucl. Phys. {\bf A579} (1994) 453;\\  D.V. Bugg. Nucl. Phys. {\bf B88} (1975) 381\\ [ 5 pt ]
2. B.M.Abramov et al. Nucl. Phys. {\bf A542} (1992) 579\\ [ 5 pt ]
3. A.E.L.Dieperink, P.K.A.de Witt Huberts. Ann. Rev. Nucl. Part. Science {\bf 40} (1990) 239 ( see also references in this review ) \\ [ 5 pt ]
4. T.W.Donnelly, J.D.Walecka. Ann. Rev. Nucl. Science {\bf 25} (1975) 329\\ [ 5 pt ]
5. E.J.Moniz. Phys. Rev. {\bf 184} (1969) 1154
\newpage
\begin{tabular}{cc}
\epsfxsize=70mm \epsfbox{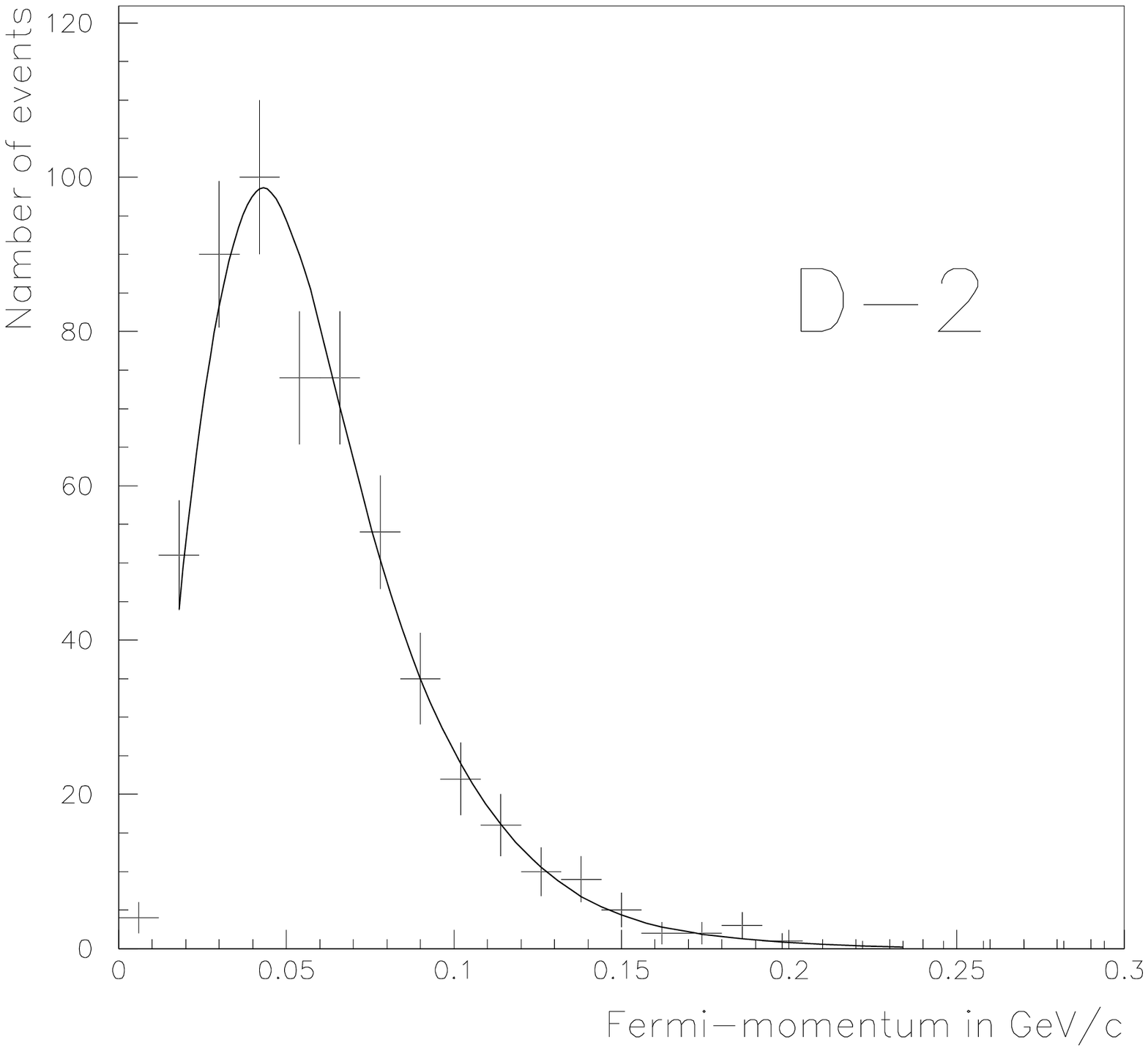} & \epsfxsize=70mm \epsfbox{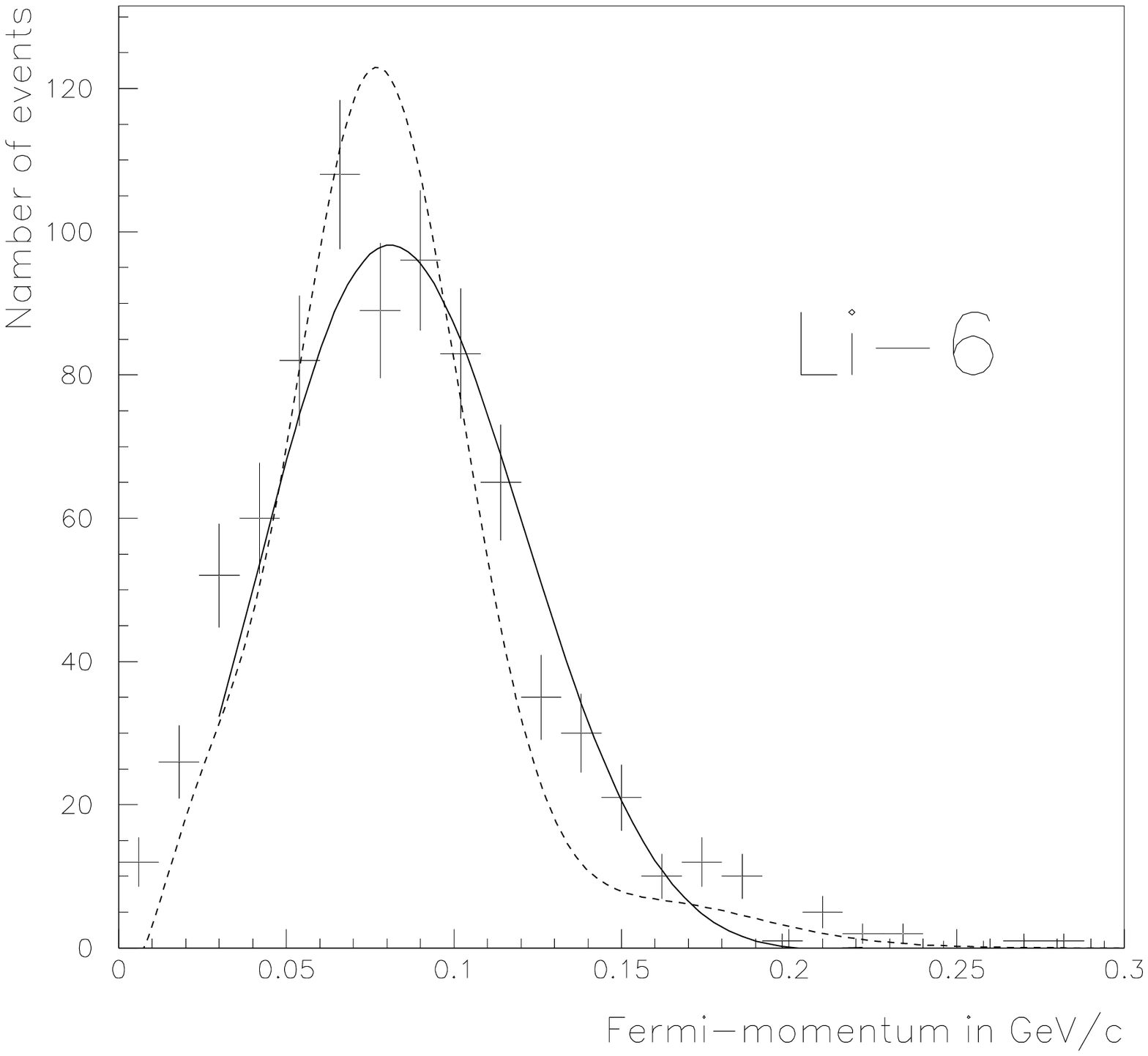} \\
\epsfxsize=70mm \epsfbox{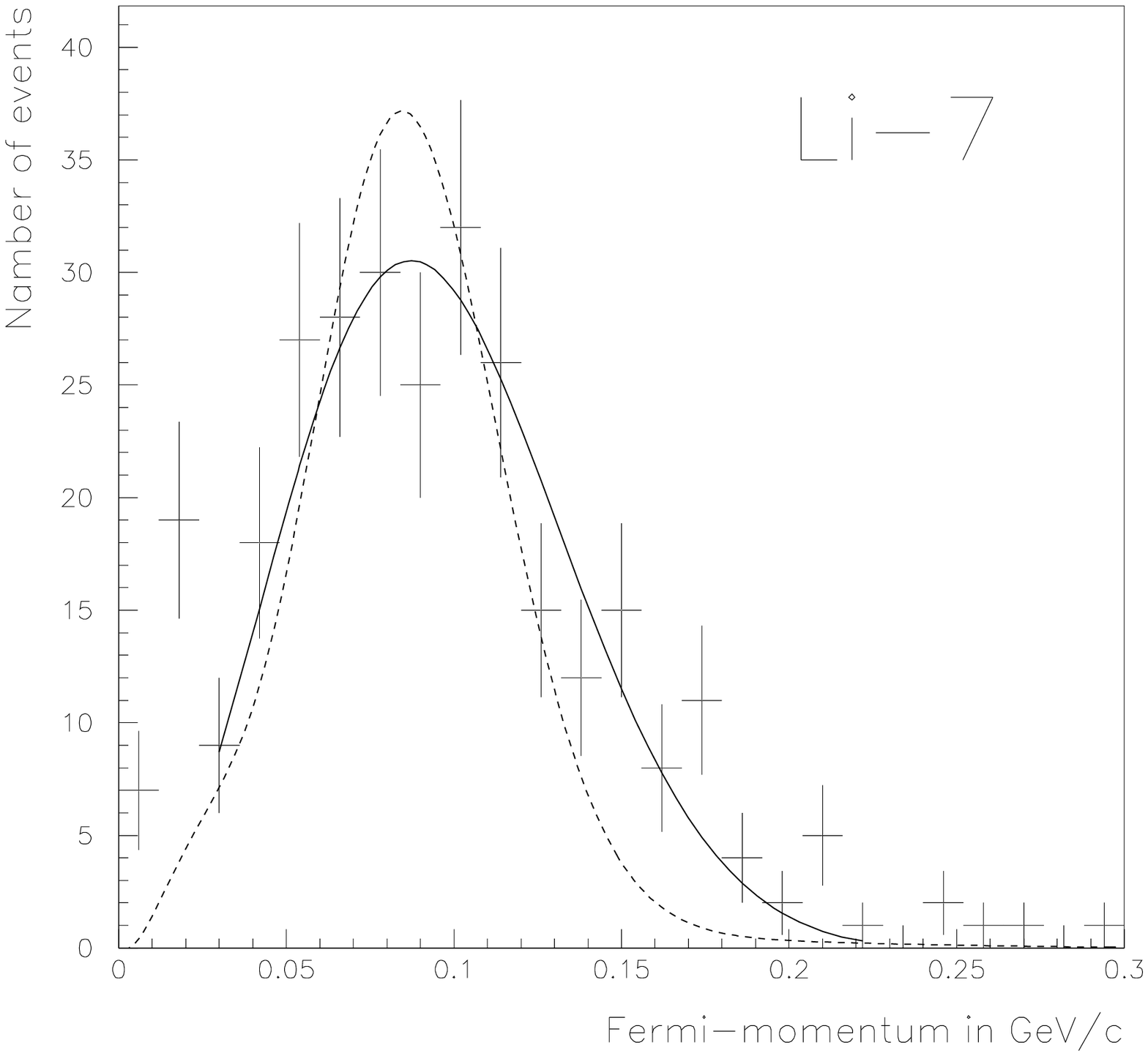} & \epsfxsize=70mm \epsfbox{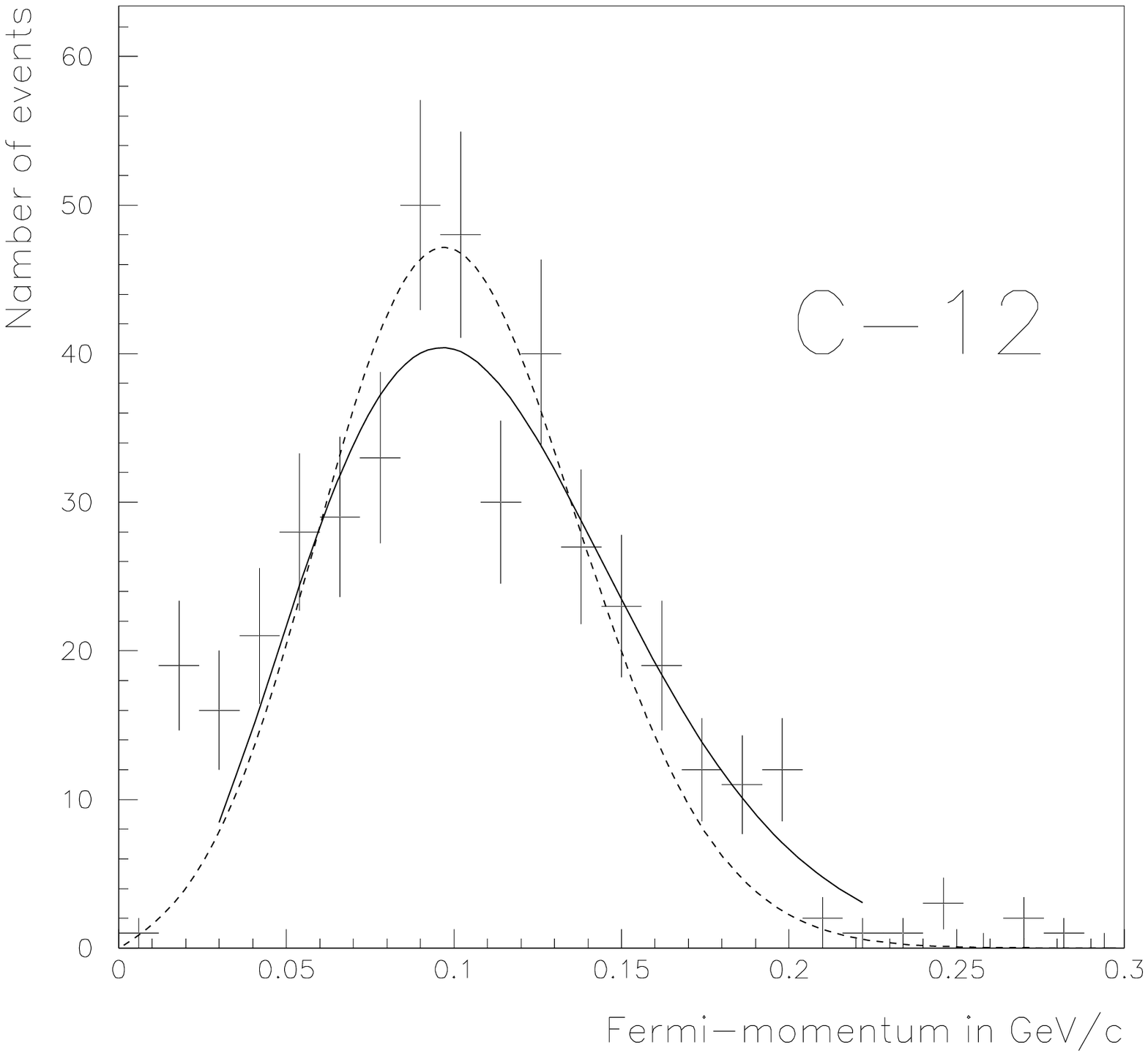} \\
\end{tabular}
\begin{center}
Fig. 1.  Fermi-momentum distributions for the $^2D$, $^6Li$, $^7Li$ and $^{12}C$ nuclei at p$_0 =$ 0.7 GeV/c  and their description with Hulten wave function for $^2D$, in the Fermi-gas model (dashed curve) and in oscillator model (solid curves) for the other nuclei.
\end{center}
\newpage
\begin{flushleft}
Table 1 \\
Parameters of  the Fermi-momentum distributions  in MeV/c.\\
\end{flushleft}
\begin{tabular}{|c|c|c|c|}
\hline
 Nucleus & This experiment & Other measurements & Wave function\\
\hline
$^2D$ & $\beta = 264 \pm 44$  & 260 & Hulten \\
\hline
$^6Li$ & $k_F = 96 \pm 16,0$  & $169 \pm 5$ [4] & Fermi-gas  \\
       & $k_0 = 73 \pm  1,0$  &           & Oscillator \\
\hline
$^7Li$ & $k_F = 119 \pm 6,0$  &  & Fermi-gas \\
       & $k_0 = 89 \pm 2,1$  &  & Oscillator \\
\hline
$^{12}C$ & $k_F = 137 \pm 3,2$  & $221\pm5$ [4] & Fermi-gas \\
       & $k_0 = 99 \pm 2,1$  & 131 [5] & Oscillator  \\
\hline
\end{tabular}
\end{document}